\documentclass[aps,preprint,groupedaddress]{revtex4}

\newcommand{\invdetst}{\frac{\Lambda}{\alpha \sqrt \gamma}} 
\newcommand{\mnras}{{\it Monthly Not.\ Roy.\ Astron.\ Soc.}} 

\newtheorem{lemma}{Lemma}

\begin{document} 

\draft 
\title{A Class of Exact Solutions to the Blandford-Znajek Process} 
\author{Govind Menon} 
\address{Dept.\ of Mathematics and Physics,\\
Troy University, Troy, Alabama, 36082} 
\author{Charles D.\ Dermer} 
\address{E. O. Hulburt Center for Space Research, Code 7653, 
Naval Research Laboratory, Washington, DC 20375-5352} 
\date{\today} 
\begin{abstract} 
We analyze the constraint equation giving allowed solutions 
describing fields and currents in a force-free magnetosphere 
around a rotating black hole. Utilizing the divergence properties of the 
energy and angular-momentum fluxes for physically allowed solutions, 
we conclude that poloidal surfaces are independent of the radial 
coordinate for large values of $r$. Using this fact and the Znajek
regularity condition, we 
explicitly derive all possible exact solutions admitted by the constraint 
equation for $r$ independent poloidal surfaces, which are given in terms of
the electromagnetic angular velocity function $\Omega = 1/a\sin^2\theta$,
where $a$ is the angular 
momentum per unit mass of the black hole. 
\end{abstract} 


\maketitle


Blandford and Znajek \cite{bz77} 
proposed a mechanism whereby the rotational energy of 
black holes could be extracted through electromagnetic 
processes. In this model, the black-hole magnetosphere is 
force-free, and the currents and fields are determined 
self-consistently in the Kerr geometry. They derived a constraint 
equation for the functions governing the system, and then 
imposed a regularity condition \cite{zna77} at the black-hole 
event horizon. 

The general relativistic approach of Blandford and Znajek \cite{bz77} 
was recast in terms of a  $3+1$ absolute space and global time formalism 
by Thorne and collaborators \cite{mdt82,tmd82,tpm86} and, more recently, 
by Komissarov \cite{kom04}. Using the approach of Ref.\ \cite{kom04}, 
we \cite{md05} rederived the Blandford and Znajek constraint equation in the $3+1$ formalism, from 
which far-field solutions were then derived that matched the Znajek regularity 
condition. Here we derive the first exact class of solutions for the constraint 
equation. 

Our absolute space is described by a surface of constant time $t$ in the Boyer-Lindquist coordinates 
$(t,r,\theta,\varphi)$ of the Kerr geometry, with a metric of the form 
\begin{equation} 
ds^2=(\beta^2-\alpha^2)dt^2+2\beta_\varphi d\varphi dt + \gamma_{rr} dr^2 + \gamma_{\theta\theta} d\theta^2 
+ \gamma_{\varphi\varphi}d\varphi^2 
\label{ds2} 
\end{equation} 
(see Refs.\cite{kom04,md05} for the values of the metric coefficients and a fuller discussion 
about the derivation of some the results used here). 

In a force-free situation, the electric field $E$ is transverse to the magnetic field $B$, 
so that $E\cdot B = 0.$ Stationarity and axisymmetry imply that the toroidal component of 
the electric field, $E_\varphi = 0$. Consequently there exists a vector 
$\omega = \Omega \partial_\varphi$ such that  $E = -\, \omega \times B$. 
Here $\Omega$ is the angular velocity of the electromagnetic field, and surfaces 
of constant $\Omega$ define poloidal surfaces. 
The constitutive equations 
relating the electromagnetic field and its dual for materials with zero dielectric and magnetic 
susceptibilities are 
\begin{equation} 
E = \alpha D + \beta \times B\;, \;{\rm and} \;H = \alpha B - \beta \times D\;. 
\label{suscept1} 
\end{equation} 
Because $B$ has zero divergence, the poloidal $(r,\theta)$ component of $B$ 
can be written as 
\begin{equation} 
B_P = \frac{\Lambda}{\sqrt\gamma}(-\Omega_{,\theta}  \partial_r + \Omega_{,r} 
\partial_\theta)\;. 
\label{bpexplicit} 
\end{equation} 
Here $\gamma = \det(\gamma_{ij})$, and $\Lambda$ is an arbitrary function that is constant on poloidal surfaces (such 
functions are called poloidal functions). 

The electric charge density $\rho$ is given by 
\begin{equation} 
\sqrt \gamma \rho = \partial_r[\invdetst (\gamma_{\varphi \varphi} \Omega +\beta_\varphi) 
\gamma_{\theta \theta} \Omega_{,r}]+\partial_\theta 
[\invdetst (\gamma_{\varphi \varphi}\Omega +\beta_\varphi)\gamma_{r r} \Omega_{,\theta}]. 
\label{rhoformula} 
\end{equation} 
The toroidal component $J^\varphi$ of the electric current density vector is given by 
\begin{equation} 
\sqrt \gamma J^\varphi = 
\partial_r[\invdetst (\alpha^2-\beta^2-\beta_\varphi \Omega) 
\gamma_{\theta \theta} \Omega_{,r}]+ \partial_\theta 
[\invdetst (\alpha^2-\beta^2-\beta_\varphi \Omega)\gamma_{r r} \Omega_{,\theta}]. 
\label{jphiformula} 
\end{equation} 
As can be seen, the quantities $\rho$ and $J^\varphi$ are uniquely described by the poloidal functions $\Omega$ 
and $\Lambda$ and the metric coefficients. 
The vanishing of the curl of $E$ means that 
$B \cdot \nabla H_\varphi =0$, and therefore, that $H_\varphi $ is poloidal and therefore a function of $\Omega$ alone. 
We write the constraint equation as 
\begin{equation} 
\frac{1}{2 \Lambda } \frac{d H_\varphi^2}{d \Omega}=\alpha (\rho \Omega \gamma_{\varphi \varphi}-J_\varphi). 
\label{finalconsb} 
\end{equation} 
The physical meaning of eq.\ (\ref{finalconsb}) is that when $\Omega$ and $\Lambda$ are correctly 
picked, the right hand side of the above equation is a function of $\Omega$ alone, 
so that it is possible to integrate and obtain an expression for $H_\varphi$. 
It is important to realize that when $\Omega$, $\Lambda$, and $H_\varphi$ are fixed, the 
fields and currents are uniquely prescribed. 

Given the fields and currents, the flux of energy is given by 
the Poynting vector $S = E\times H$, which is divergence-free for 
a force-free, time-independent system. The radial Poynting flux $S^r = - \Omega H_\varphi B^r$. 
Likewise, the angular-momentum flux vector is divergence-free, and the radial 
angular-momentum flux $L^r = - H_\varphi B^r$. 
The net rates of energy and angular-momentum extraction from a rotating black hole 
are given, respectively, by 
\begin{equation} 
\frac{d \cal E}{dt}=\int S^r \sqrt {\gamma_{rr}} dA=-\int H_\varphi \Omega B^r \sqrt {\gamma_{rr}} dA\;, 
\label{engext} 
\end{equation} 
and 
\begin{equation} 
\frac{d \cal L}{dt}=\int L^r \sqrt {\gamma_{rr}} dA=-\int H_\varphi  B^r \sqrt {\gamma_{rr}} dA\;. 
\label{angext} 
\end{equation} 

Since the above two expressions only differ by the
presence of $\Omega$ in the right hand side, as
in \cite{md05}, we conclude that $\Omega$ is
asymptotically $r$ independent. With this in mind,
for the remainder of this paper, we will assume that
$\Omega_{,\; r}=0$, as an initial step of research into
the search for exact solutions. Consequently, $\Omega$
is a function of $\theta$ alone, and 
poloidal surfaces are surfaces of constant $\theta$. 
Inserting Eqs.\ (\ref{rhoformula}) and (\ref{jphiformula})
into Eq.\ (\ref{finalconsb}) 
for our $r$-independent $\Omega$, the 
constraint equation reduces to the form 
\begin{equation} 
\frac{1}{2 \Lambda } \frac{d H_\varphi^2}{d \Omega} =  \frac{\alpha \gamma_{\varphi \varphi}}{\sqrt\gamma} 
[\Omega \partial_\theta(\invdetst (\gamma_{\varphi\varphi} \Omega+ \beta_\varphi) \gamma_{rr} \Omega_{,\theta}) 
+ \partial_\theta(\invdetst (\beta^2 - \alpha^2+\beta_\varphi \Omega)\gamma_{rr} \Omega_{,\theta})]\;. 
\label{finalconsc} 
\end{equation} 
For a consistent formulation of the theory of 
axisymmetric, stationary, force-free magnetospheres, the above assumption implies that $H_\varphi$ and $\Lambda$ are 
to be a function of $\theta$ alone.    
For the case of a Kerr black hole in Boyer-Lindquist coordinates, the constraint equation becomes 
\begin{equation} 
\frac{1}{2 \Lambda } \frac{d H_\varphi^2}{d \Omega} = \frac{\sin\theta}{\rho^2} 
[\Omega \partial_\theta(\frac{\Lambda \Omega_{,\theta}}{\sin\theta} (\gamma_{\varphi\varphi} \Omega+ \beta_\varphi) ) 
+ \partial_\theta(\frac{\Lambda \Omega_{,\theta}}{\sin\theta}(\beta^2 - \alpha^2+\beta_\varphi \Omega))]\;. 
\label{finalconsc2} 
\end{equation} 
Here $\rho^2 = r^2+a^2 \cos^2\theta$. 
Expanding the above equation to order $(1/r^3)$ \cite{md05} gives 
$$-\frac{1}{2 f(\theta)} \frac{d H_\varphi^2}{d \theta}=-\Omega \sin\theta \frac{d}{d\theta}(f\Omega \sin\theta) 
+\frac{\sin\theta}{r^2}[-a^2\Omega\sin^2\theta\frac{d}{d\theta}(f\Omega \sin\theta)+\frac{d}{d\theta}(\frac{f}{\sin\theta})] $$ 
\begin{equation} 
+2M\frac{\sin\theta}{r^3}[a\Omega\frac{d}{d\theta}(f \sin\theta(1-a\Omega\sin^2\theta))-\frac{d}{d\theta}(\frac{f}{\sin\theta}(1-a\Omega\sin^2\theta))]\;, 
\label{assym} 
\end{equation} 
where $f(\theta)\equiv -\Lambda \Omega_{,\; \theta}\equiv A_{\varphi,\; \theta}$. 
Since poloidal functions are function of $\theta$ alone, all the terms proportional to the inverse powers of $r$ 
must vanish identically for appropriate choices of $f$ and $\Omega$ since $H_\varphi$ is to be a poloidal function. Consequently, only the zero$^{th}$-order term survives, implying that 
\begin{equation} 
H_\varphi^2= \pm H_0 ^2+(f\Omega \sin\theta)^2. 
\label{hphi} 
\end{equation} 
The choice of $\Omega$ is determined by the Znajek \cite{zna77} regularity 
condition, which is given by 
\begin{equation} 
H_\varphi=\frac{\sin\theta}{\rho_+ ^2} (2r_+ M \Omega-a)f, 
\label{EHBC} 
\end{equation} 
where the subscript $+$ indicates that the relevant quantities are to be evaluated at the event horizon and $\rho_+^2 = r_+^2 + a^2 \cos^2\theta$. 
From Eqs.\ (\ref{hphi}) and (\ref{EHBC}), we can eliminate $H_\varphi$ and find the relation between $f$ and $\Omega$. Explicitly, 
\begin{equation} 
\pm H_0 ^2=\frac{\sin^2\theta}{\rho_+ ^4} [(4r_+^2 M^2-\rho_+ ^4) \Omega^2-4r_+ M a\Omega+ a^2]f^2. 
\label{nonzeroho} 
\end{equation} 

\begin{lemma} 
If $\Omega \neq 1/a \sin^2\theta$, then $f$ is given by the expression 
\begin{equation} 
f = \frac{B_0 \sin\theta}{\sqrt{\mid (a\Omega\sin^2\theta)^2 -1 \mid}}\,, 
\label{feq} 
\end{equation} 
where $B_0$ is a constant. 
\end {lemma} 
{\bf Proof 1.} If $f$ is to be a solution to Eq.\ (\ref{finalconsc}) for a given form of $\Omega$, 
then the pair $(f,\Omega)$ 
should remove all the $r$-dependence in Eq.\ (\ref{assym}). In particular, the vanishing of the $1/r^2$ term implies that   
\begin{equation} 
a^2\Omega\sin^2\theta\frac{d}{d\theta}(f\Omega \sin\theta)=\frac{d}{d\theta}(\frac{f}{\sin\theta}). 
\label{secondorder} 
\end{equation} 
As was shown in Ref.\ \cite{md05}, Eq.\ (\ref{feq}) is the unique solution to 
the above equation when $\Omega \neq 1/a \sin^2\theta$. $\S$ 
\vskip0.25in 

Eqs.\ (\ref{nonzeroho}) and (\ref{feq}) can be used to completely determine all the 
possible allowable forms of $\Omega$ (when $\Omega \neq 1/a \sin^2\theta$). 
To this end define 
\begin{equation} 
\Omega_\pm = \frac{a}{2 M r_+ \pm \rho_+ ^2}, 
\end{equation} 
noting that $\Omega_- = 1/a \sin^2\theta$. 

Therefore, a necessary condition that the pair $(f,\Omega)$ would generate a self consistent 
solution to the stationary, axis-symmetric, force-free solution for a magnetosphere in the Kerr geometry is that they 
satisfy Eq.\ (\ref{nonzeroho}) and Eq.\ (\ref{feq}) for fields to be regular at the event horizon (as long as $\Omega \neq 1/a \sin^2\theta$ ). 

\begin{lemma} 
If $\Omega \neq 1/a \sin^2\theta$, the only choices for $\Omega$ are 
\begin{equation} 
\Omega = \tilde{\Omega}_\pm =\frac{\tilde{A}\Omega_+ \pm \tilde{B} \Omega_-}{\tilde{A}\mp \tilde{B}} 
\label{badsol} 
\end{equation} 
where $\tilde{A}=B_0^2\neq 0$ (if $B_0 = 0$, the fields are trivial), and $\tilde{B}=H_0^2 \rho_+^4 \Omega_+ \Omega_-$. 
\end {lemma} 
{\bf Proof 2.} 
From Eq.(\ref{nonzeroho}) we see that 
\begin{equation} 
f^2 = \frac{\pm H_0^2 \rho_+^4}{a}\frac{\Omega_+ \Omega_-^2}{(\Omega-\Omega_+)(\Omega-\Omega_-)}. 
\end{equation} 
Here the $\pm$ factor is to ensure that $f^2 \geq 0$. Similarly we find from Eq.\ (\ref{feq}) that 
\begin{equation} 
f^2 = \frac{B_0^2}{a^2 \sin^2\theta \mid(\Omega-\Omega_-)(\Omega+\Omega_-)\mid }. 
\label{fp} 
\end{equation} 

Equating the right-hand sides of the last two equations, we see that 
\begin{equation} 
B_0^2  \mid \Omega-\Omega_+\mid =H_0^2 \rho_+^4 \Omega_+\Omega_- 
\mid \Omega+\Omega_- \mid. 
\end{equation} 
The above equation has the unique solution given by Eq.\ (\ref{badsol}). $\S$ 
\vskip0.25in 
In the event that $H_0 \rightarrow 0$, we see that $\tilde{\Omega}_\pm \rightarrow \Omega_+$, so that 
the $\Omega_-$ solution is never realized by $\tilde\Omega_\pm$ since $B_0 \neq 0$. 

\begin{lemma} 
No solutions exist to the constraint equation that satisfy the Znajek event horizon regularity 
condition when $\Omega=\tilde{\Omega}_\pm$. 
\end{lemma} 
{\bf Proof 3.} 
If $\Omega=\tilde{\Omega}_\pm$ satisfies the constraint equation (to all orders in $r$), 
then $f$ is given by 
Eq.\ (\ref{feq}). It is then sufficient to show that $\Omega=\tilde{\Omega}_\pm$, 
along with $f$ as given in Eq.\ (\ref{feq}) does not satisfy Eq.\ (\ref{assym}) to order $1/r^3$. 
Vanishing of the $1/r^3$ term in Eq.\ (\ref{assym}) implies that 
\begin{equation} 
\frac{d g^2}{d\theta} \sin\theta (1-a \tilde{\Omega}_\pm \sin^2\theta)= 2 g^2 \cos\theta(a \tilde{\Omega}_\pm \sin^2\theta+1), 
\label{thirdorder} 
\end{equation} 
where $g=f(1-a \tilde{\Omega}_\pm \sin^2\theta)$. Inserting the expression for 
$f$, and $\tilde{\Omega}_\pm$ in the definition of $g$, we find that 
\begin{equation} 
g^2=\frac{\pm 1}{r_+^2 + a^2}(H_0^2 \rho_+^4-B_0^2 \rho_+^2 \sin^2\theta)\,, 
\label{gsquare} 
\end {equation} 
where $\pm$ ensures that $g^2 > 0$. Inserting Eq.\ (\ref{gsquare}) in Eq.\ (\ref{thirdorder}) we find that 
\begin{equation} 
-\frac{2 M r_+B_0^2}{a^2}=\sin^4\theta[B_0^2 a^2 \sin^2\theta-B_0^2\rho_+^2-2H_0^2 a^2 \rho_+^2]. 
\end{equation} 
The above equation will not be satisfied because the left hand side is independent of $\theta$ unlike the 
right hand side. Therefore, we reach a contradiction. $\S$. 

\begin{lemma} 
$\Omega = \Omega_-$ is an exact solution to the constraint equation 
(Eq.\ (\ref{finalconsc2})) where $\Lambda$ is any arbitrary poloidal function. 
\end{lemma} 
{\bf Proof 4.} 
To simplify the discussion, we first make the following observations 
$$\beta^2-\alpha^2+2\beta_{\varphi}\Omega_-+\gamma_{\varphi \varphi}\Omega_-^2=\frac{\rho^2}{a^2 \sin^2\theta}$$ 
$$\beta^2-\alpha^2+\beta_{\varphi}\Omega_- = -1$$ 
\begin{equation} 
\gamma_{\varphi \varphi} \Omega_-+ \beta_{\varphi}=\frac{r^2+a^2}{a}\,. 
\end{equation} 
The constraint equation (Eq.\ (\ref{finalconsc2})) can be rewritten as 
$$\frac{1}{2 \Lambda } \frac{d H_\varphi^2}{d \Omega}=\sin\theta(\frac{\Lambda \Omega_{-,\theta}}{\sin\theta})_{,\theta} 
\frac{\beta^2-\alpha^2+2\beta_{\varphi}\Omega_-+\gamma_{\varphi \varphi}\Omega_-^2}{\rho^2}+$$ 
\begin{equation} 
\Lambda \Omega_-\Omega_{-,\theta}\frac{(\gamma_{\varphi \varphi} \Omega_- + \beta_{\varphi})_{,\theta}}{\rho^2}+\Lambda \Omega_{-,\theta}\frac{(\beta^2-\alpha^2+\beta_{\varphi}\Omega_-)_{,\theta}}{\rho^2} 
\end{equation} 
$$=\Omega_-(\frac{\Lambda \Omega_{-,\theta}}{a \sin\theta})_{,\theta}$$ 
The right hand side is clearly a poloidal function, thus removing the only 
constraint for arbitrary values of $\Lambda$. $\S$ 
\vskip0.25in 

It is easily checked that $\Omega_-$ satisfies the regularity condition when $H_0 = 0$,
and when $H_\varphi = + f\Omega \sin\theta$. 
Therefore, from the above lemma, we see that 
\begin{equation} 
H_\varphi=\frac{2}{a^2} \Lambda \frac{\cos\theta}{\sin^4\theta}. 
\label{finalhphi}
\end{equation} 
If we put all the lemmas together, we get our main result: \\ 
{\bf Theorem} 
{\it When poloidal surfaces are surfaces of constant $\theta$, the unique class of solutions to the stationary, axisymmetric force-free 
magnetosphere that is regular on the event horizon of a Kerr black hole is generated 
by the function $\Omega=\Omega_-$. The entire degree of freedom 
in this theory lies in the poloidal but otherwise arbitrary function $\Lambda$. 
} 

The importance of this result is that all exact solutions 
to the Blandford-Znajek process when $\Omega_{,\; r}=0$ are constructed from the 
various choices of $\Lambda$ and $\Omega_-$, with $\Lambda$ chosen
to give physically allowed expressions for fields, charges, and
current densities along the poles.
\\
{\bf Corollary}
{\it It is impossible to extract energy (and angular momentum) from a stationary,
axisymmetric force-free magnetosphere (that is regular on the event horizon)  of a Kerr black hole when $\Omega_{,\; r}=0$.}

{\bf Proof.} From Eq.\ (\ref{bpexplicit}), Eq.\ (\ref{engext}), Eq.\ (\ref{angext}) and
Eq.\ (\ref{finalhphi}) we see that

\begin{equation} 
\frac{d \cal E}{dt}=-\frac{8\pi}{a^4} \int_{0}^{\pi} \frac{\Lambda^2 \cos^2\theta}{\sin^9\theta}d\theta \leq 0 \;,  
\end{equation} 
and 
\begin{equation} 
\frac{d \cal L}{dt}=-\frac{8\pi}{a^3} \int_{0}^{\pi} \frac{\Lambda^2 \cos^2\theta}{\sin^7\theta}d\theta \leq 0\;.\S  
\end{equation}

The $\Omega_+$ solution of Ref.\ \cite{md05} is an approximate 
solution that may be realized as the far-field limit of an exact
solution only if poloidal functions become $r$ dependent (as we
get closer to the event horizon). This solution has the nice feature
that it yields positive energy extraction that is in accord
with results of numerical simulations \cite{mck05,kra05} that
utilize magnetic fields sustained 
by external accretion disks. 
As such, the inability to extract energy from a black hole
for the set of exact solutions we have derived indicates that
the condition $\Omega_{,\; r}=0$ must necessarily be relaxed.

\vskip0.5in 
G.\ M.\ acknowledges funding through a Troy University sabbatical. 
This research is also supported 
through NASA {\it GLAST} Science Investigation No.\ DPR-S-1563-Y. 
The work of C.\ D.\ D.\ is supported by the Office of Naval Research.

\end{document}